# Absolute measurement of the nitrogen fluorescence yield in air between 300 and 430 nm.

G. Lefeuvre[a][1], P. Gorodetzky[a][2], J. Dolbeau[a], T. Patzak[a], P. Salin[a]

[a] *APC - AstroParticule et Cosmologie, CNRS : UMR7164 – CEA – IN2P3 – Observatoire de Paris, Université Denis Diderot - Paris VII, 10, rue Alice Domon et Léonie Duquet 75205 Paris Cedex 13, France*

*Abstract*

The nitrogen fluorescence induced in air is used to detect ultra-high energy cosmic rays and to measure their energy. The precise knowledge of the absolute fluorescence yield is the key quantity to improve the accuracy on the cosmic ray energy. The total yield has been measured in dry air using a $^{90}$Sr source and a [300-430 nm] filter. The fluorescence yield in air is

$$4.23 \pm 0.20 \text{ photons per meter}$$

when normalized to 760 mmHg, 15°C and with an electron energy of 0.85 MeV. This result is consistent with previous experiments made at various energies, but with an accuracy improved by a factor of about 3. For the first time, the absolute continuous spectrum of nitrogen excited by $^{90}$Sr electrons has also been measured with a spectrometer. Details of this experiment are given in one of the author's PhD thesis [32].

*Keywords:* nitrogen fluorescence, air fluorescence, extensive air showers, ultra-high energy cosmic rays.

PACS: 29.30.Dn ; 98.70.Sa

---

[1] Actual address: Department of Physics, Syracuse University, Syracuse, NY 13244, USA
[2] Corresponding author: *E-mail address:* philippe.gorodetzky@cern.ch

**Introduction**

One of the current challenges in high energy particle physics is to find the origin and nature of Ultra-High Energy Cosmic Rays (UHECR, E > $10^{18}$ eV). The measurement of their energy spectrum could confirm their interaction with the cosmological background (GZK theory). But this is a difficult task: the UHECR are not detectable themselves, but only the air shower they induce while going through the atmosphere.

Experiments which have tried to solve this problem still have non consistent results for E > 5·$10^{19}$ eV. They use different detecting techniques. On one hand, AGASA [1, 2] has an array of ground detectors separated by about 0.7 km. Muons and electrons of the showers arriving at the ground are sampled. The energy reconstruction is difficult because of the lack of knowledge on hadronic cross sections. This method relays on Monte-Carlo particle distributions in the shower approximation to find the energy of the incoming particle. On the other hand, HiRes [3] uses fluorescence telescopes to detect the continuous development of the shower. The accuracy of this method can still be improved if the uncertainties on the fluorescence itself are lowered. Today, the Auger experiment uses both methods but still calibrates energy extracted from the ground detectors by the fluorescence (also on the ground, [4]). In the future, space-based experiments like JEM-EUSO will look for the fluorescence signal from above [5].

Since 1964, several authors [6, 7, 8] have measured the fluorescence yield of each emission band of nitrogen (see the spectrum in fig. 5). The fluorescence yield is defined as the number of photons produced when an electron goes through one meter of air. In 2002, the cosmic ray community [9, 10, 11] decided to start an active campaign of fluorescence measurements for cosmic ray physics. Some were made at high energy in electron accelerators, the energy ranging from 80 MeV to 28 GeV [12, 13]. Others were made at low energy, around 1 MeV with a $^{90}$Sr radioactive source, or much lower with electron guns ([14, 15, 18, 21]). These experiments confirmed the hypothesis that the fluorescence yield, up to some tens of MeV, is proportional to the energy deposit of the particle, dE/dX. The difference between energy loss and energy deposit is of no importance from 0.1 to 10 MeV [16, 17], which includes our range. Therefore, the absolute scale of the yield can be set with a measurement performed with electrons from a $^{90}$Sr source.

Now, the relative variation of the fluorescence yield with altitude is generally considered as quite well known [18]. The parameterization of the yield as a function of altitude is deduced from the relationship between pressure and altitude in an atmospheric model [19, 20]. These parameterizations are fairly consistent with one another, but can only give the variation of the yield with altitude and not the absolute value of the yield. Moreover, measurements of the absolute fluorescence yield have never reached a precision better than 13 to 20 % [13, 21].

The experiment presented in this paper has been realized with a double purpose:
- to improve significantly the precision on the fluorescence yield: it is now 5 % ;
- to achieve a first continuous measurement of the fluorescence spectrum in the range from 300 to 430 nm, which had never been done before with a radioactive source.

**1. Experiment**

*1.1. Principle*

Two measurements have been carried out, named "integral" and "spectral" measurements. Electrons from a strontium source are used to excite nitrogen. Three detectors are needed, all of them based on Photonis photomultiplier tubes (PMT). One detects the electrons (electron-PMT), the other two, the photons (photon-PMTs).

On one hand, the nitrogen emits fluorescence light through bands emitting lines from 300 to 430 nm. Thus cosmic ray experiments like HiRes and Auger use band pass filters of that range in front of their detectors. This is what is done here in the integral measurement.

On the other hand, fluorescence yield experiments generally use narrow filters to separate the lines. But the overlap between some spectral filter bandwidth and the close proximity of some nitrogen bands make the separation difficult in reality. For this reason, the spectral measurement has been performed with an optical grating spectrometer, allowing a continuous check and a good separation of the lines.

*1.2. Description of the setup*

This setup has been conceived with the aim of improving the measurement accuracy. The fluorescence chamber is a cross-shaped stainless steel chamber (a 6-ports accelerator pipe cross). It is schematically shown in fig. 1. Electrons are emitted by a strong $^{90}$Sr source (370 MBq) which is placed on the top part of the cross. On the bottom part, a plastic scintillator (truncated cone, 20 and 28 mm diameters, 20 mm thick and its angle is the same that the one made by the electron trajectories at the cone periphery). The electrons path is hence defined as the inside of a cone (100 mm height), with the source at its top. To homogenize the light between the scintillator and the PMT, a kaleidoscopic hexagonal light guide is used to carry the light. The phototube is a Photonis XP2262 [22] equipped with an active voltage divider (the rate is high, of the order of 2.5 MHz, and very stable).

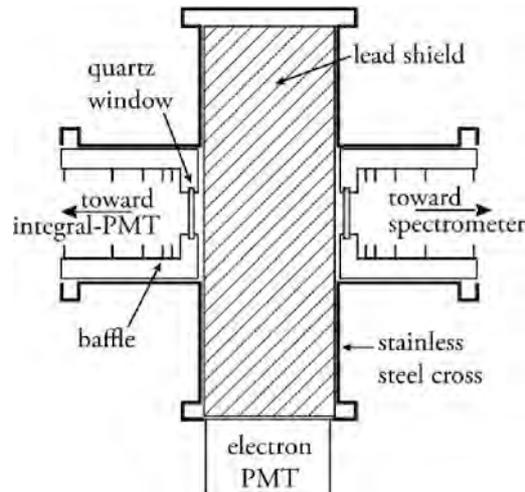

Figure 1. Schematic view of the fluorescence chamber. The strontium source and the scintillator are *inside* the lead shield. The internal structure of the lead is shown in fig. 2 (side view).

The consequence of the high counting rate is a high X-ray noise level, due to the interacting electrons with the surrounding matter, including the PMT. To shield to a maximum these X-rays, a 10 cm diameter lead cylinder has been placed inside the chamber with the $^{90}$Sr source at its centre. A vertical cone is dug in the lead so the electrons can reach the scintillator. This cone is 30 mm diameter at its base, slightly wider than the electrons cone defined by the scintillator. Such a cone minimises the number of interactions between electrons and the lead: either they are totally stopped (around the source), or they touch in a grazing way the cone in their way to the scintillator. A GEANT Monte-Carlo simulation shows that half of the useful electrons do reach the scintillator without an interaction with the lead, the other half touching only once.

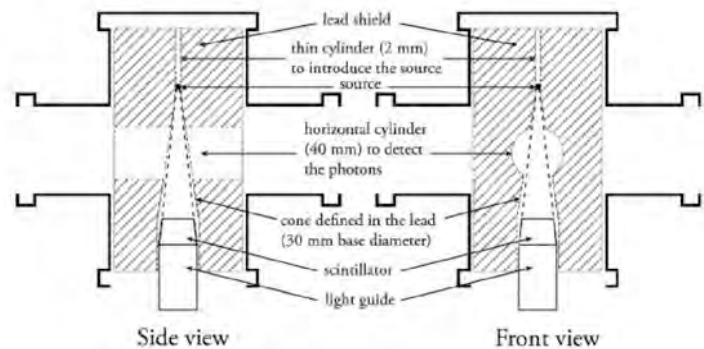

Figure 2. Internal structure of the lead shield. Inside the hollow cone is the smaller cone of the path of the useful electrons defined by the source and the scintillator (thick dotted line).

An horizontal cylinder (40 mm diameter) is also dug in the lead, centred on the optical axis, to let the photons go to the photon-PMTs. This structure is represented in fig. 2.

The effective fluorescence volume is thus, on a first approximation, the intersection of the vertical cone with the horizontal cylinder. In a more precise way, all geometrical efficiencies have been determined with Monte-Carlo simulations. They take into account all possible interactions between electrons and their surroundings and give the exact photon solid angles as well as the exact useful electron path length (the length used in the yield expressed in photons per meter).

Two kinds of interactions can occur: with the gas and with the lead. First, part of the energy of the source electrons is lost by ionization, i.e. creation of secondary electrons. If the secondaries go too far away from the useful fluorescence volume, they can either reach the lead or enter the horizontal cylinder. Both modify the detection efficiencies (geometry wise) of the photon-PMTs calculated. But 99 % of these secondaries have an energy smaller than 5 keV. Their range in air at atmospheric pressure is less than 2 mm. This has been checked through a geometrical simulation not to be enough to have an influence on the solid angles. The effect of these δ rays on the yield expressed in photons per deposited energy is explained later in paragraph 3.1

The second type of interaction concerns the electrons from the source that do not reach directly the scintillator. Some of them are scattered by the lead and then bounce back to the scintillator. This has three consequences, studied by a GEANT Monte-Carlo simulation. The first is the increase in the electron counting rate, due to the larger solid angle available for the electrons: it is doubled by this effect. The second is a modification in the electron energy spectrum, which contains more low energy electrons. This has been taken into account in the calculation of the electrons average energies. The third is a slight change in the useful electron path length, also taken into account.

Some electrons can also bounce from the scintillator back into the fluorescence volume. This has been simulated to be a $10^{-9}$ effect on the photon detection, hence totally negligible.

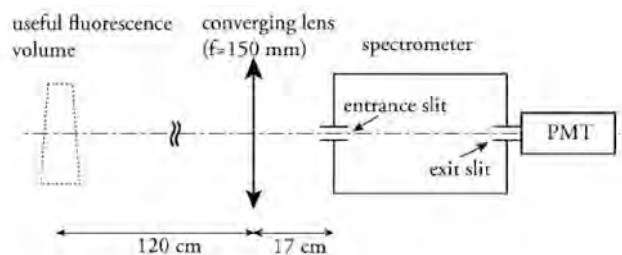

Figure 3. Schematic view of the optical components for the spectral measurement.

For the integral measurement, fluorescence photons are detected and counted by a Photonis XP2020Q (2 in. diameter) placed on the left hand-side of the setup [23]. This PMT will be named integral-PMT and it is equipped with a fused silica window. As the fluorescence efficiency is known to be low (around 4 photons per meter), it has to work in single photoelectron mode, meaning at a very high gain. It is polarized positively (photocathode at ground) in order to reduce its own noise level to around 300 Hz (instead of 3 kHz with a negative polarity). This effectively removes the tiny discharges in the silica between the photocathode and the outside world. The optical filter on the entrance window is a Schott-Desag BG3 (2 mm thick, 34 mm diameter) [24]. It is the same as JEM-EUSO intends to use, and very similar to those used by Auger. It is glued with Epotec 301-2 [25] which has the same refraction index than both the window and the filter.

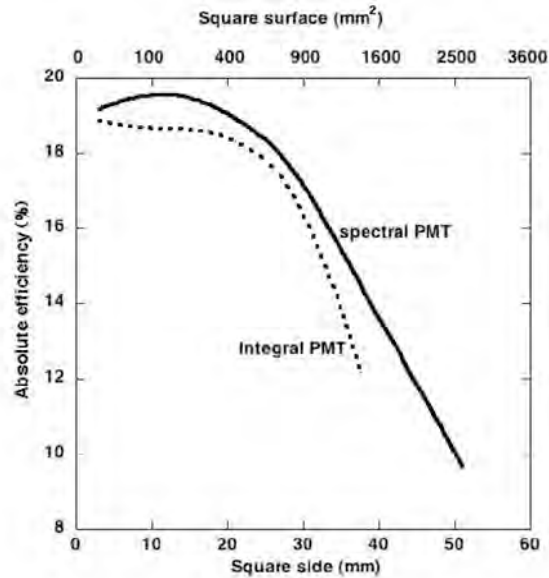

Figure 4. Absolute efficiency of the integral-PMT (dashed line) and spectral-PMT (bold line) as a function of the size of the effective detecting surface size. A 20 mm diameter diaphragm is used to limit the effective detecting area of the integral-PMT to the flattest region of the photocathode. Using a diaphragm of 20 mm diameter (314 mm$^2$), the efficiency is 18.7%. The spectral-PMT is used in its central region (~ 10×18 mm) where the efficiency including the filter is 19.3%. The uncertainty is 1.7% of the efficiency values. Note that the zero suppression on the vertical axis.

For the spectral measurement, photons are analyzed by an optical grating spectrometer (Jobin-Yvon H25) and counted by another Photonis XP2020Q, attached to the spectrometer. This PMT will be named spectral-PMT. The light rays incoming the spectrometer have to be inside its numerical aperture (f = 250 mm, NA = f/4), so a silica converging lens of 150 mm focal length and 46 mm diameter [26] is inserted at the right place between the fluorescence volume and the entrance slit of the spectrometer. The optical image of this volume is thus on the latter. The optical magnification is 1/7. Entrance and exit slits of the spectrometer have the same dimensions: 2 x 7 mm, the maximum size available, inducing a resolution (measured with lasers) of 6 nm FWHM. This part of the setup is shown in fig. 3.

The main uncertainty in this kind of experiments, where very few photons are counted, arises from the absolute efficiency of the photon-PMTs themselves. The manufacturers provides this value with an uncertainty of 15 to 20 % (at 1σ), which is not precise enough. For this reason, we designed a new way to measure the absolute efficiency of the photon-PMTs in the single photoelectron mode very accurately, based on the comparison with a NIST photodiode.

The relative efficiency map of the photocathodes of the PMTs working in single photoelectron mode was measured with a 377 nm LED every 3 mm in X and Y with a relative precision better than 0.5%. This precision was needed to be better than the point to point variations in efficiency (about 2%), in order to control these variations. A measurement of the absolute efficiency of one of the map points with an accuracy of 1.7%, transforms that relative efficiency map in an absolute efficiency map. A dedicated paper, following a patent, will present this measurement. This result was used to determine the flattest region of the photocathode (see fig. 4). The effective detecting area of the integral-PMT was limited by a 20 mm diameter diaphragm where the efficiency stays roughly constant when the diaphragm size is increased. That way, a small error on the diaphragm diameter has a negligible consequence on the PMT efficiency (it has however on the solid angle, but this is easy to control).

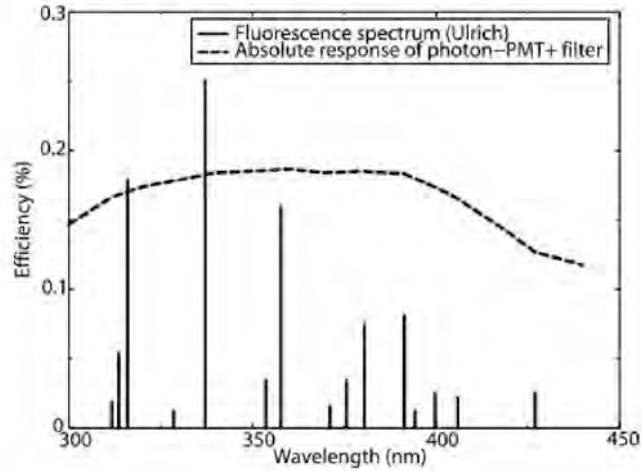

Figure 5. Fluorescence spectrum (vertical lines) with relative intensities and absolute response of the photon detector (PMT + filter + diaphragm, dashed curve) (source of the fluorescence spectrum : [14]). The integral-PMT efficiency for this spectrum is 17.8%.

The absolute spectral efficiency of the detector {PMT + filter + diaphragm} is the dashed line superimposed on the fluorescence spectrum in fig. 5. Its value for photons at 377 nm is $(18.9 \pm 0.3)\%$, and, when convoluted with the fluorescence bands [14] and the relative variations given by Photonis [22], is $(17.8 \pm 0.4)\%$. The spectral-PMT is illuminated on a 10 x 18 mm area (smaller area than the 20 mm circle for the integral PMT, hence all photons at the exit slit of the spectrometer reaching different parts of this area would here also have roughly the same probability to produce photoelectrons).

On their way from the fluorescence chamber to any of the photon-PMTs, integral or spectral-PMT, the 20 cm long tubes are baffled to prevent reflected photons to reach the photocathode.

*1.3. Gas*

The gas used is either nitrogen for the adjustments or Messer dry air for the measurements. Fig. 6 shows the gas setup. It is possible to mix gases and to introduce controlled quantities of different impurities and water vapour. Those measurements including pressure variations are planned for the future.

The gas circulates at a rate of 1 $L.h^{-1}$ through the chamber to avoid impurities build-up and degassing of the walls. This circulation is controlled by a precise flow meter [27], followed by a pressure controller [28] and an ultra-clean primary pump with an ultimate vacuum of 50 mbar [29]. Internal and external pressures and temperatures are regularly monitored during data taking.

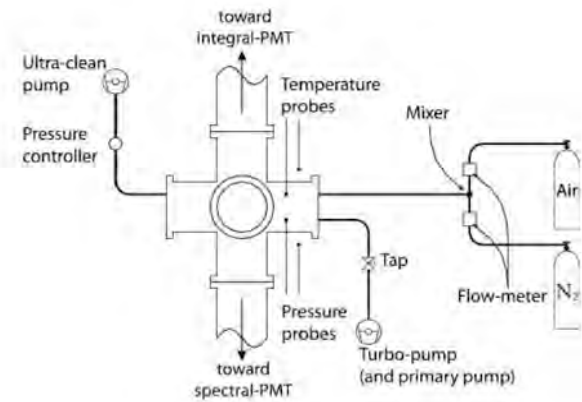

Figure 6. The gas circulation scheme.

*1.4. Data acquisition*

Hardware data acquisition is basically the same that for other authors, but had to be adapted to high counting rates. Fig. 7 shows its principle.

The lifetimes of excited levels of nitrogen molecules at atmospheric pressure are of the order of a few nanoseconds [8, 14, 15, 18, 21]. A measurement based on the time coincidence between an electron and a photon is made to discriminate fluorescence from background photons.

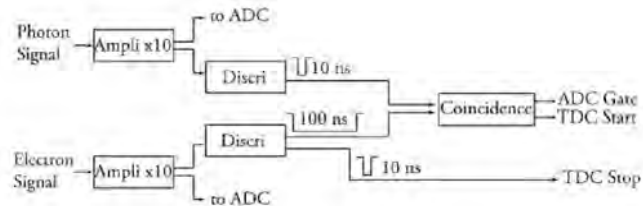

Figure 7. The data acquisition principle

Basically, an electron in the scintillator is the trigger for the photon measurements. An electron produces about 0.16 photon in the 4 cm long fluorescence volume. Geometrical efficiencies toward each photon-PMT are very small: $3.69 \cdot 10^{-4}$ for the integral-PMT and $7.48 \cdot 10^{-6}$ for the spectral-PMT. As a consequence, only $5.9 \cdot 10^{-5}$ and $1.2 \cdot 10^{-6}$ photon respectively reach the PMTs. So, both photon-PMTs work in single photoelectron mode. Their spectra are very "pure" in "1 photo-electron", with a negligible amount of "2". So a discriminator is enough to select the "1 photoelectron" peak. This can be seen in fig. 8.

ADCs (LRS 2249A, 0.25 pC per channel) record the electron and photon spectra. They are used only to check the gain stability of all PMTs during data taking. The "detection inefficiency", D, defined as the proportion of lost fluorescence counts due to the mandatory discriminator threshold on the single photoelectron spectrum (see fig. 8), has also been measured and been found to be 3.76 % of the measured single photoelectron peak.

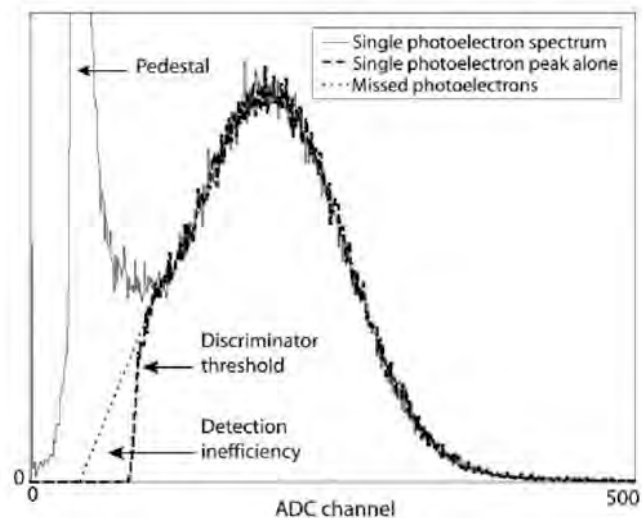

Figure 8. Single photoelectron spectrum, with and without the pedestal.

Time to Digital Converters (TDCs) (LRS 2228A, 0.3 ps per channel, monohit) are used to record the time difference spectra between electrons and photons.

Both ADCs and TDCs could be triggered by the electron signal, but this would produce a very high random rate. We cleaned up the trigger by replacing it by an electron-photon coincidence.

This coincidence is performed by a fast NIM coincidence unit. The photon pulse is 10 ns wide. The electron pulse is 100 ns wide to take into account the delay in photon emission (lifetimes of the levels). This was set to prepare acquisitions at low pressure, where the lifetimes are longer than at atmospheric pressure. Even at very low pressure, 100 ns ensures the loss due to that effect will be negligible. In other terms, a coincidence occurs when the photon arrives with less than 100 ns delay after the electron.

The TDC is stopped by the same electron (see fig. 7), this time as a 10 ns wide pulse and delayed by 150 ns. Therefore, the start has a rate much lower than the stop and this inverts the time axis. Results can be seen in fig. 9. The useful signal is the peak in the spectrum. It can be easily separated from the flat background. The effect on the peak left slope due to the lifetimes of the nitrogen levels cannot be seen on this figure because it has been measured in air at atmospheric pressure. The narrow peak on the far right-hand side is due to random photons arriving some nanoseconds before electrons. These spectra suffer from an important dead time. But the ratio of the peak to total spectrum is not affected. The real number of counts is extracted from scalers measuring the total spectrum with a small and well corrected dead time. These fast scalers (CAMAC CERNSPEC NE003, 25 MHz, and VME V560E, 100 MHz) count all electron, photon and coincidence pulses.

Two thresholds, creating two triggers, are set on the $^{90}$Sr spectrum to study the possible dependence of the fluorescence yield with the electron energy even if it is difficult to imagine an influence other than energy deposit. One is set at ~ 600 keV and another at ~ 1.2 MeV. The corresponding average energies are 1.1 and 1.5 MeV. There are around 1 000 photon pulses per second counted by the integral-PMT (with a background of 300 Hz) and 40 by the spectral-PMT on at the top of the 337 nm line (with a background of 30 Hz). Moreover, the dead times, $T_M$, of both electron scalers have been specifically measured and found to be 1.2 % (scaler rate $2 \cdot 10^6$ Hz) and 0.7 % (scaler rate $6 \cdot 10^5$ Hz).

## 2. Results

### 2.1. Integral measurement

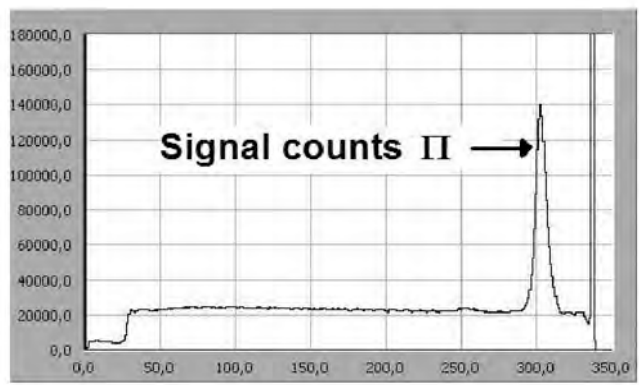

Figure 9. Example of TDC spectrum for the integral measurement

The photon yield per unit length, Y, is derived from the signal portion of the TDC spectrum through the following formula:

$$Y = \frac{\Pi \cdot \left(\frac{C}{H}\right) \cdot (1+D)}{N_e (1+T_M) L_{moy} \, \varepsilon_{PMT} \, T_W \, \varepsilon_{geo}}$$

where:
- $\Pi$ is the number of signal counts in the TDC spectrum (see fig. 9) ;
- H is the integral of this spectrum ;
- C is the number of coincidences sent to the TDC measured by a scaler. Only H coincidences have effectively started the TDC due to its dead time ;
- 1 + D is the correction to the detection inefficiency ;
- $N_e$ is the number of electrons according to the scaler ;
- $1 + T_M$ is the correction to the dead time of the scaler itself ;
- $L_{moy}$ is the mean length of the electrons path in the fluorescence volume ;
- $\varepsilon_{PMT}$ is the efficiency of the integral-PMT ;
- $T_w$ is the transmission of the fused silica window closing the fluorescence chamber ;
- $\varepsilon_{geo}$ is the geometrical efficiency.

Systematic errors are presented in table 1. The main uncertainty of this experiment is due to the high counting rate of the electrons, leading to a non-linear dead time dependence in the TDC module. This effect, which varies from channel to channel, depends on the internal TDC time constants and is not fully understood. It has been evaluated by using all the module's channels, and another TDC module (CAEN V1290N, multihit) to compare their results. All values are found equal within 4% (at 1$\sigma$) and this uncertainty has been chosen in a conservative way. In the future, to further reduce this uncertainty, fast flash ADCs will be used, allowing to discriminate pulses with and without pile-up on an event per event basis.

On the contrary, the uncertainty on the efficiency of the integral-PMT is very low. This is due to the photomultiplier efficiency measurement made especially for the fluorescence measurement by the patented method.

| Quantity | Uncertainty |
|---|---|
| 1+D, $1+T_M$ | negligible |
| geometry ($L_{moy}$, $\varepsilon_{geo}$) | 1.5 % |
| $T_w$ | 1 % |
| $\varepsilon_{PMT}$ | 1.7 % |
| TDC | 4 % |
| *TOTAL* | 4.7 % |

Table 1. Systematic uncertainties of the experiment.

An example of typical data sample is presented in tab. 2. The low (1.1 MeV) and high (1.5 MeV) energy measurements give respectively 3.95 and 4.34 photons per meter. Statistical uncertainties are 0.2 % and 0.8 %.

The energy normalization is made at 0.85 MeV using the dE/dX ratios with values interpolated from NIST data [30] and yields respectively 4.05 and 4.41 photons per meter.

| Gas | dry air |
|---|---|
| Pressure | 753.8 mmHG |
| Temperature | 22.8 °C |
| Mean energy | 1.1 MeV |
| Duration | 66 582 s |
| Counts in signal portion | 1 597 715 |
| Signal counts $\Pi$ | 1 039 798 |
| TDC integral H | 8 833 521 |
| Coincident counts C | 12 680 216 |
| Electron counts $N_e$ | $1.286 \cdot 10^{11}$ |
| Fluorescence yield Y | 3.95 photons/m |

Table 2. Data sample for the low energy measurement.

The pressure and temperature normalizations are then applied. Each excited level has its own lifetime, and the yield can be written with respect to pressure and temperature using the kinetic theory [6]. Thus this normalization has been made for each band using previous parameterizations (which give the same variations for the first kilometers above the ground [20]). Here, Nagano's model [21], who uses the different bands yields is used. The normalized pressure and temperature are those of the US-Standard 1976 model [19] at sea level: 760 mmHg and 15°C. In this model, for an electron energy of 0.85 MeV, one finds that the ratio of the yield at 753.8 mmHg and 295.95° K (the 1.1 MeV conditions) to the yield at 760.0 mmHg and 288.15° K (the US Standard conditions), is 0.9863, and the similar ratio for the 1.5 MeV conditions which are 751.8 mmHg and 296.05° K is 0.9860. These ratios are then used to normalize the measured values.

Finally, the two normalized values are 4.05 and 4.42 ph/m and have, due to this normalization, an added uncertainty of 0.6 %, setting the total relative uncertainty to 5.0 %. They are separated by 8.5%, inside an error bar (± 1σ). Hence, these are averaged to get the fluorescence yield, at 760 mmHg, 15°C and for 0.85 MeV electrons:

$$4.23 \pm 0.21 \text{ photons / m.}$$

This yield per meter is that of the primary particle, therefore it takes into account its production of δ rays and their fluorescence. This number can also be written in units of photons per deposited energy. For a given energy, these two quantities are strictly proportional, since:

$$Y_m = \sum_\nu \left(\frac{\phi_\nu}{h\nu}\right) \rho \frac{dE}{dX}$$

where $\Phi_\nu$ is the fluorescence efficiency at the wavelength corresponding to the frequency ν, with the yield per deposited energy being [31]:

$$Y_E = \sum_\nu \left(\frac{\phi_\nu}{h\nu}\right)$$

The energy of the δ rays is naturally included in the dE/dX of the initial electron. Nevertheless, in the context of this experiment, a special care must be taken for the δ rays. A Monte-Carlo simulation has been performed and shows that here, with the geometry described earlier and at atmospheric pressure, only 1 % of the δ rays have more than 5 keV and some of these could be lost in the lead before having produced any fluorescence. The effective dE/dX should thus be reduced. This simulation shows then that almost 67 % of these δ rays with 5keV or more will produce detectable fluorescence. Therefore only 33 % of them are effectively lost and contribute to the reduction of the dE/dX. Finally, this reduction, hence the reduction in yield, is 0.4 %. The central value is shifted from 0.4 % (we measured 4.21 photons/m before applying this correction). The uncertainty stays the same at 5 %. This would be very different at lower pressures, where a more extensive Monte-Carlo simulation would be required and the correction would be increased.

The energy deposited by a 0.85 MeV electron in a meter of the US Standard air is 0.2059 MeV [30]. The fluorescence yield per deposited energy is thus $20.38 \pm 0.98$ photons per MeV.

The nitrogen yield to air yield ratio has also been measured, (at one electron energy only: 1.1 MeV), and found to be $4.90 \pm 0.01$, where the low uncertainty is only statistical: all systematic errors compensate, the gas being the only element changed from one measurement to the other. This ratio is compatible with what has been measured at different energies : $6 \pm 2$ at 28.5 GeV [13], $5.5 \pm 0.3$ at 1 MeV [21]

*2.2. Spectral measurement*

Spectral measurements are interesting for many purposes. The gas kinetic theory is valid for individual bands. Is the sum of the different bands yield equal to the integrated yield used by the majority of cosmic ray experiments using fluorescence? How do the different bands change yield when temperature and pressure are modified? Up to now, the different bands have been observed either with a grating spectrometer [6, 7, 14] in a first method, the electrons being produced by an electron gun. Their energy (around 10 keV) is so small that the electron scattering in the gas prevents any measurement of path length, hence cannot give a yield in photons per meter. The second method [8,15, 18, 21] uses narrow interference filters to analyze the bands. Here, the electrons, as in this work, come from a $^{90}$Sr source. Fig. 1 of [21] illustrates the method complexity. It is difficult to separate overlapping bands in some filters. The only solution to have a good resolution on the bands is to use a grating spectrometer, and the easiest way to have electrons with an energy large enough to know their path is to use a $^{90}$Sr source. The acceptance of a spectrometer is roughly an order of magnitude smaller than that of a filter. So the very first step is to prove the feasibility of such a method. This is what is done here.

Hence, in this experiment, a $^{90}$Sr source 10 times stronger than in [8,15, 18, 21] was used, The counterpart of the high number of incident electrons is that a very special attention has to be given to pile-up effects and to the lead shield geometry to avoid the introduction of a large X-ray background. Results of the spectral measurements are represented in fig. 10. The whole fluorescence spectrum has been measured, from 300 to 436 nm, in dry air at room temperature and normal pressure, with electrons of average energy of 1.1 MeV. The time allowed to measure this spectrum was short and the spectrometer equipped with an output slit can measure only one wavelength at a time. So it was decided to open the slits at their maximum, setting the 1σ resolution to 3 nm. Thus measurements were made with steps of 3 nm. But even with $2 \cdot 10^6$ electrons detected every second, only 0.16 fluorescence photons per second are recorded in the signal part of the TDC spectrum for the most intense line (337 nm). This is the reason why this spectrum has only been made for the "low" energy threshold, as defined above.

Fluorescence lines are much narrower than the resolution set for the spectrometer as is seen in [14]. The absolute yield is thus given by the height of the curve at a given wavelength, and not by the area integrated over a bandwidth dλ. The sum of the yields of all lines indicated in fig. 10 is $3.9 \pm 0.8$ photons per meter. This value is in close agreement with the result of the integral measurement. The 20% uncertainty is due most entirely to the bad knowledge of the spectrometer absolute efficiency, especially in the UV.

Superimposed on this spectrum is Ulrich's spectrum [14], convoluted with a 3 nm resolution. As it is a relative yield spectrum, it is shown here normalized to the 337 nm line. Our spectrum is well compatible with what would measure [14] with such a resolution.

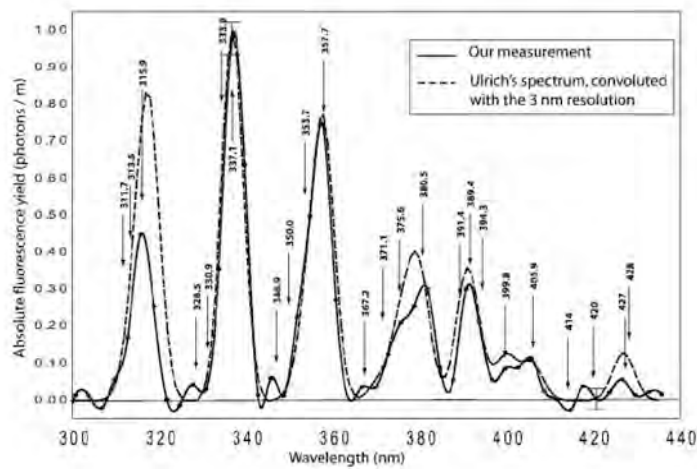

Figure 10. Spectrum of the absolute fluorescence yield of nitrogen in air between 300 and 436 nm. The bold line is the result of the present experiment, and the dashed line is Ulrich's spectrum [14] as if analyzed by our spectrometer. Discrepancies could be explained by the ageing of our spectrometer.

The first observation is the confirmation of the feasibility of this measurement: all the main lines are indeed observed. Evidence is given that an absolute spectrum can be measured with a basic apparatus, and taking care of reducing the PMT background as explained earlier. The sum of the lines yield is consistent with the previous integral measurement. Moreover, its uncertainty is 20 %, when previous experiments do not give better than 15 %.

The second observation concerns the discrepancies, around 316 nm and beyond 375 nm. The extraction of the absolute yield involves the efficiency of the spectrometer. Its spectral efficiency curve is provided by the manufacturer with a low accuracy in the UV (which is true also for Ulrich's results). Furthermore two effects induce a very important loss in our spectrometer:
- the ageing of each optical element (mirrors and grating). The spectrometer efficiency can be reduced to a value around 15 % after ten years of use ;
- the extensive use with intense UV light before this experiment.

At 400 nm, the absolute efficiency of the spectrometer was measured with an accuracy of about 2% to be only 15 %, instead of the 61 % given by the manufacturer (who confirms in a private communication such a low value compatible with ageing). The method used is very similar to that taken to determine the absolute efficiency of the photon-PMTs (comparison to a NIST photodiode). This value of 15 % has been used to calculate the absolute yields in the entire spectral range. But there is no reason that this loss is constant with respect to wavelength. The 20% uncertainty on the yield arises from this unknown but limited variation. The question of getting the absolute efficiency of spectrometers in the UV is quite challenging and is the object of specific attentions by the community of atomic / spectral physics.

It is unfortunate that this spectrometer method does not yield yet an accuracy better than the "integrated yield" method. It will if the spectrometer resolution can be made high enough to totally separate the bands, which is possible if it is equipped with a CCD readout to minimize the experiment duration. Then, the absolute efficiency of the spectrometer will have to be determined with a high accuracy in the UV, not an easy task according to Ulrich, but possible through our patented method.

In the future, two measurements will be done:
- calibrate in an absolute way the old Jobin-Yvon;
- use the new spectrometer able to measure the full spectrum at once with a 0.1 nm resolution and equipped with a light intensifier to measure the fluorescence yield.

3. Conclusion

The absolute fluorescence yield of nitrogen in dry air at atmospheric pressure has been measured. The precision of the measurement is improved by a factor of three, which has an immediate impact on the cosmic ray energies found by HiRes which uses Bunner's [6] yield. Their energies are increased by 22%, hence their spectrum is much closer to AGASA's (which incidentally have been recently lowered the energy of their points by 10% [33]).

The first continuous fluorescence spectrum of nitrogen excited by electrons from a $^{90}$Sr source was also measured.

Next steps are to introduce impurities in the gas, such as argon, water vapour and pollutants. A pressure study of the total yield will be made. On another hand, the spectral measurement will be improved thanks to a new spectrometer. Papers will follow to account for these measurements, which will provide an overall and realistic view of the fluorescence phenomenon.

**Acknowledgements**

We would like to thank Bernard LEFIEVRE for his help with GEANT simulation of the setup, and François LELONG and Jean-Paul RENY for their help in building the bench.